\documentclass[twocolumn,english]{revtex4-1}
\usepackage[T1]{fontenc}
\usepackage[latin9]{inputenc}
\usepackage{array}
\usepackage{color}
\usepackage{float}
\usepackage{textcomp}
\usepackage{multirow}
\usepackage{tipa}
\usepackage{tipx}
\usepackage{amstext}
\usepackage{graphicx}
\usepackage{natbib}

\makeatletter

\newcommand*\LyXbar{\rule[0.585ex]{1.2em}{0.25pt}}
\providecommand{\tabularnewline}{\\}

\usepackage{chngcntr}

\makeatother

\usepackage{babel}
\begin{document}

\title{Ab-initio Study of the Trimethylaluminum Atomic Layer Deposition
Process on Carbon Nanotubes \textemdash{} An Alternative Initial Step}

\author{Anja Förster $^{\ddagger,\dagger,\bullet}$}

\author{Christian Wagner$^{\mathsection}$ }

\author{Jörg Schuster$^{\dagger}$}

\email{joerg.schuster@enas.fraunhofer.de}

\author{Joachim Friedrich$^{\ddagger}$}

\affiliation{$\ddagger$ Technische Universität Chemnitz, 09126 Chemnitz, Germany}

\affiliation{$\dagger$ Fraunhofer Institute for Electronic Nano Systems (ENAS),
Chemnitz, Germany}

\affiliation{$\mathsection$ Center for Microtechnologies (ZfM), TU Chemnitz,
Chemnitz, Germany}

\affiliation{$\bullet$ now at: TU Dresden, Center for Advancing Electronics Dresden
(cfaed), 01062 Dresden}
\begin{abstract}
Electronic applications of carbon nanotubes (CNTs) require the deposition
of dielectric films on the tubes while conserving their excellent
electronic properties. In our density functional theory study we use
the trimethylaluminum (TMA) atomic layer deposition (ALD) as a model
process for a CNT-functionalization. Since this functionalization
relies on the presence of OH-groups, the CNTs are exposed to a water
or oxygen pre-treatment. We show that only CNTs with a single-vacancy
defect are able to react with H$_{2}$O or O$_{2}$. Further, the
defect is preferably saturated by oxygen. This leaves the CNT without
the necessary hydroxyl groups for the first TMA addition. Therefore,
we propose an alternative initial step after which a classical TMA
ALD process can be performed on the CNT.
\end{abstract}
\maketitle
\setlength\emergencystretch{1em}

\section{Introduction }

Carbon nanotubes (CNTs) are characterized by an outstanding mechanical
strength in combination with a high charge carrier mobility. The latter
property makes them promising candidates for novel nano scale transistors,
especially for high frequency applications \citep{Steiner2012,Schroeter2013},
interconnects \citep{Fiedler2014,Fiedler2015} or optoelectronic devices
\citep{Rauhut2012,Scarselli2012,Avouris2008,Blaudeck2015}. Their
conductance is also very sensitive to external influences such as
mechanical strain \citep{Yang_2000,Kleiner_2001,Wagner2012,Wagner2016}
or changes in their chemical environment promoting sensor applications
as well \citep{Lefebvre2004,Ohno2006,Sgobba2009,Scarselli2012,Blaudeck2015}.
The electronic properties of carbon nanotubes are closely related
to their structure. Depending on their chiral angle, nanotubes can
be metallic, semiconducting or semimetallic \citep{Anantram2006}.
However, any application of CNTs relies on well defined electronic
properties and thus type selected CNTs are required. Despite of ongoing
research, there is lack of type selected material today. This
is due low yield in complicated fabrication setups \citep{Rao_2012,Yang_2014}
or expensive chirality sorting procedures after fabrication \citep{Green_2011,Liu_2013,Tanaka_2015}.
Thus, tuning the electronic properties of CNTs by deposition of functional
thin films is seen as a promising approach. In addition, many device
architectures based on carbon nanotubes rely on the possibility to
deposit high quality dielectric films on CNTs, i.e. as gate isolation.

From a chemical perspective, carbon nanotubes are nearly perfect,
rolled up graphene sheets. Graphene is known to be chemically inert,
but CNTs show some reactivity which increases with curvature. This
is due to a release of stress energy during the reaction increasing
with CNT curvature. Although there is a finite reactivity, any chemical
modification of CNTs is challenging. This applies for the deposition
of functional films on CNTs by chemical vapor deposition as well.

Deposition of thin films by atomic layer deposition (ALD) has been
proven to yield uniform and conformal high quality films with a precise
thickness control on many substrates as well as on nanostructures.
In particular, ALD of Al$_{2}$O$_{3}$ based on trimethylaluminum
(TMA) and water is known to be a very robust process which is well
studied in literature \citep{ALD_George,TMA_Energy}. Like many ALD
methods the TMA ALD process relies on the existence of functional
groups on the initial surface before the first ALD cycle can be performed.
Specifically, hydroxyl groups must be present on the CNT which is
the crux of the problem.

Different approaches for the TMA ALD on CNTs have
been suggested in literature \citep{lee20032,herrmann2005multilayer,farmer2006atomic,zhan2008atomic,cavanagh2009atomic}.
For example, Zhan et al. reported in their experimental study \citep{zhan2008atomic}
that a pre-treatment with ethanol and sodium dodecylsulfate increases
the ratio of hydroxylated surfacegroups. Consequently, this pre-treatment
leads to a more conformal coating in comparison to the CNT being dispersed
in water only. Farmer and Gordon \citep{farmer2006atomic} suggested
an alternative way of performing a TMA ALD on CNTs which relies on
nitrogen dioxide (NO$_{2}$) being evenly adsorbed on the CNT\textquoteright s
surface. In the following step TMA molecules react with the oxygen
atoms of NO$_{2}$, forming a uniform layer over the CNT. While effective,
both methods introduces other atomic species such as sulfur and nitrogen.

Therefore, in our density functional theory (DFT) study we analyze
an alternative pre-treatment of CNTs in a water and oxygen atmosphere
that does not introduce any additional atomic species. Our pre-treatment
requires the presence of single vacancy (SV) defects, which make the
CNT more reactive \citep{Kozlowska2016}. This defect, characterized
by a missing carbon atom, often occurs in chemical vapor deposition
(CVD) grown CNTs \citep{Rao2015}. According to \citep{mawhinney2000surface}
a defect rate up to 5\% can be observed. We study the functionalization
of metallic single wall (5,5)-CNTs via TMA ALD. Quantum chemical methods
are used to obtain a thorough understanding of the CNT pre-treatments
required prior to ALD-functionalization as well as the relevant process
steps. In the second part of our study, the reaction pathways of the
first steps of TMA-ALD growth on the hydroxylized CNTs are analyzed
and compared to similar reactions on flat substrates. We successfully
show that TMA ALD performs well on pre-functionalized CNTs without
introducing external atomic species such as nitrogen and sulfur.

\section{Model System and Computational Details}

\subsection{Model System }

Our model system is a metallic, non-periodic (5,5)-CNT consisting
of five unit cells, leading to an overall CNT length of 12.3Å. The
open ends are saturated with hydrogen atoms. Furthermore, the model
system contains one SV defect and thus its chemical structure is C$_{99}$H$_{20}$.
The presence of the SV defect turns the model CNT into an open-shell
system. The calculation of the electronic structure of open shell
systems requires multi reference methods. Such multi reference methods
are computationally very expensive and are thus not suitable for large
systems. Therefore, we avoid the open-shell-character by saturating
the SV defect with oxygen, hydrogen or hydroxyl groups during the
pre-treatment. Because of this saturation, the electronic structure
of our model systems can be investigated by DFT, which is numerically
cheaper than multi reference methods. From the technological perspective
this approach is justified in any case as an unsaturated defect will
instantly react with surrounding water or oxygen.

\begin{figure}[h]
\begin{centering}
\includegraphics{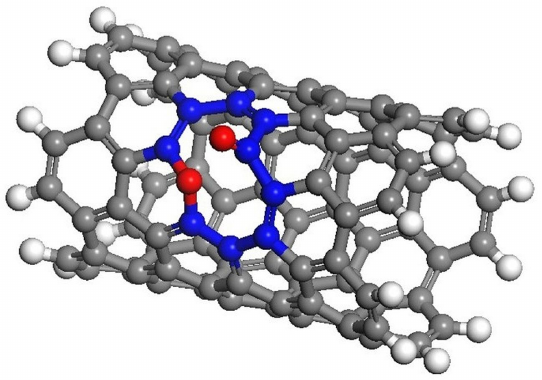}
\par\end{centering}

\caption{(Color online) Saturated CNT-structure D1 which functions as the reference
point for all calculated reaction energies of the pre-treated CNTs.
The atoms constructing the SV defect are highlighted in blue.\label{fig:1}}
\end{figure}

The avoidance of open shell systems in the electronic structure calculations
has an important consequence for calculating the reaction energies
of the pre-treatment reactions. Instead of comparing the energies
of the pre-treated CNT with the bare defective CNT, we have to choose
one of the pre-treated systems as the reference point for energy calculations.
Due to the manifold of different reaction possibilities at the CNT
defect, a systematic and intuitive labeling of the structures and
the reactions is not trivial. We decided for a naming scheme based
on the reaction type (cf. fig. \ref{fig:2}): Structures A1-I1 are
pre-treated structures (CNT + O$_{2}$), where D1 is the one with
the lowest energy and serves as a reference point for the energy calculations
\footnote{The formation energy of this structure from the CNT with the SV defect
is about -165 kcal/mol and therefore highly exothermic.} (cf. fig. \ref{fig:1}). A2-I2 are educts after wet oxygen reaction
(CNT + O$_{2}$ + H$_{2}$O), A3-I3 the educts after water reaction
(CNT + 2H$_{2}$O) and A4-L4 are structures after reaction with oxygen
and two water molecules (CNT + O$_{2}$+ 2H$_{2}$O).

According to this, reaction energies of the pre-functionalization
are calculated as:

\begin{equation}
E_{R}=E_{pre-CNT}+E_{O_{2}}-E_{CNT-D1}-E_{func}\label{eq:1}
\end{equation}

where $E_{pre-CNT}$ is the energy of the pre-functionalized CNT and
$E_{CNT-D1}$ is the energy of the functionalized reference structure
D1. $E_{O_{2}}$ is the energy of oxygen which acts as our the reference
functional group. $E_{func}$ is the energy of the functional group
of the actual functionalization reaction under consideration.

\subsection{Computational Details}

In our DFT study Acccelry\textquoteright s Materials Studio (Version
6.0) \citep{Accelrys} and Turbomole (Version 6.3) \citep{TURBOMOLE}
are used. Materials Studio is used for the calculation of the TMA
ALD process, whereas Turbomole is used for high-throughput calculations
of the CNT pre-treatment structures. In detail, the calculation setups
are the following:

For the calculations with Turbomole the BP86 functional \citep{BP_1,BP_2,BP_3,BP_4,BP_5}
together with the def2-SVP basis set for geometry optimization and
frequency analysis is used. The Resolution of Identity (RI) approximation
is applied to save calculation time \citep{TZVP_RI}. On top, higher
accuracy energy calculations with the def2-TZVP basis set \citep{TZVP,TZVP_RI}
are performed. The SCF energy convergence criterion is set to $10^{-6}$
Ha while the geometry convergence is set to $10^{-4}$ Ha for both
basis sets.

\begin{table}[h]
\begin{centering}
\caption{Overview of the calculation details. \label{tab:1}}

\par\end{centering}

\begin{centering}
\begin{tabular}{lcc}
\hline
\hline
\multirow{2}{*}{} & \multirow{2}{*}{TurboMole } & Materials \tabularnewline
 &  & Studio\tabularnewline
\hline
Functional  & BP86 & PBE\tabularnewline
Basis set (geo. & \multirow{2}{*}{Def2-SVP} & \multirow{2}{*}{DNP}\tabularnewline
 optimization) &  & \tabularnewline
Basis set (total & \multirow{2}{*}{Def2-TZVP} & \multirow{2}{*}{DNP}\tabularnewline
 energy) &  & \tabularnewline
Scf criteria & 10$^{-6}$ Ha & 10$^{-5}$ Ha\tabularnewline
Geometry opt.  & \multirow{2}{*}{10$^{-4}$ Ha} & \multirow{2}{*}{10$^{-4}$ Ha}\tabularnewline
criteria &  & \tabularnewline
Orbital cutoff & \textipa{\LyXbar{}} & 4.8 Å\tabularnewline
Smearing & \textipa{\LyXbar{}} &  $\leq$0.005 Ha\tabularnewline
\hline
\hline
\end{tabular}
\par\end{centering}

\end{table}

Within Materials Studio Dmol\textthreesuperior{} \citep{Delly_Dmol,Delly_Dmol_2}
is used for the electronic structure calculations of the molecular
systems. The functional of choice was PBE \citep{PBE} which is used
along with the DNP basis set \citep{Delly_Dmol}, whose accuracy is
similar to the def2-TZVP basis set of Turbomole \citep{Delly_PBE}.
The convergence criteria for the SCF energy calculations is set to
$10^{-5}$ Ha and $10^{-4}$ Ha for the geometry optimizations while
the orbital cutoff is fixed at 4.8 Å. Due to convergence problems
Fermi-Dirac-smearing with a maximum value of 0.005 Ha have been applied
for some calculations. For a number of reference systems, the results
of Dmol\textthreesuperior{} and Turbomole have been compared
in order to make sure that both codes give identical results within
the expected error range (cf. App. A3 of \citep{BA}). The settings
of both DFT codes are summarized in table \ref{tab:1}. For all reactions
the whole CNT is allowed to relax.

\section{Results and Discussion}

\subsection{CNT Pre-treatment }

Deposition of thin films by ALD relies on the presence of functional
groups on the surface. In the case of alumina ALD by using TMA as
a precursor, hydroxyl groups are required. Consequently, we study
the pre-treatment of defective CNTs in a reactive atmosphere consisting
of oxygen and water prior to the actual TMA-ALD process.

In detail, we analyze reactions of one SV-defect on the (5,5)-CNT
with 1) one oxygen molecule 2) one oxygen and one water molecule,
3) two water molecules and 4) with one oxygen and two water molecules
(cf. fig. \ref{fig:2}). The different reactions may result in the
CNT possessing 1) two oxygen atoms, 2) one oxygen and two hydroxyl
groups, 3) two hydroxyl groups and two hydrogen atoms or 4) four hydroxyl
groups as functional groups. The reaction of the CNT with oxygen alone
does not result in a hydroxylation. However, its energetics has to
be considered in order to determine if a hydroxylation or an oxidation
is preferable in case of concurrent reactions.

\begin{center}
\begin{figure}[h]
\includegraphics{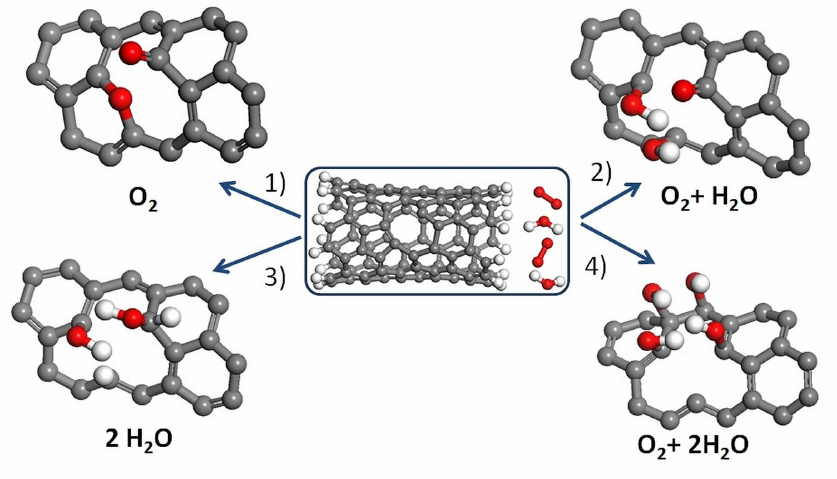}\caption{(Color online) a) Schematic illustrations of reactions 1)-4).\label{fig:2}}
\end{figure}

\par\end{center}

\begin{figure}[h]

\begin{centering}
\includegraphics{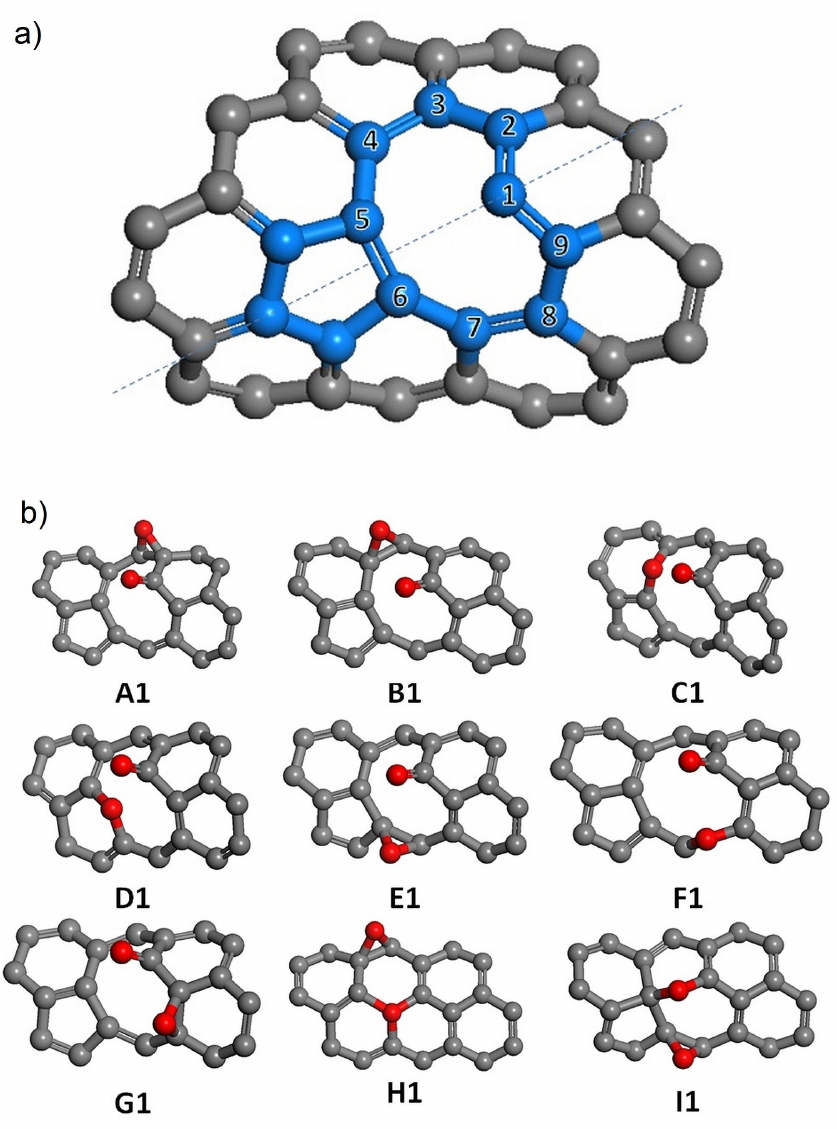}
\par\end{centering}

\caption{(Color online) a) Naming schemes of the carbon sites. The SV defect
is highlighted in blue and the dashed line displays the mirror symmetry
axis. b) Illustration of reaction 1) for the configurations A-I.\label{fig:3}}
\end{figure}

The SV-defect offers 9 positions where reactions can occur (see fig.
\ref{fig:3}a for illustration). In total, four bonds can be formed
on the defect site. All positions are likely to form a single bond,
except of position 1, where a double valence is preferred. Among the
different positions, number 1 has shown to be the most reactive during
a pre-screening of the reactive sites. Thus, we focus on reaction
configurations where position 1 is involved while the remaining sites
are varied systematically around the defect. Table \ref{tab:2} gives
an overview about the index of the structure and the according reaction
sites. Figure \ref{fig:3}b illustrates the different configurations
exemplary for reaction 1) which involves two oxygen atoms. For the
case of reaction 4) only a few typical configurations have been calculated
in our study since the number of possible configurations is very high.
Explicit structures of the correspondence of sites and functional
groups are provided in the supporting material (see figure \ref{fig:Si_1},
\ref{fig:Si_2} and \ref{fig:Si_3}).

\begin{table}[h]
\caption{Naming scheme for the configurations and the involved carbon sites
on the SV-defect, Double bonded sites are shown in bold. The carbon
sites are displayed in fig. \ref{fig:2}. \label{tab:2}}

\begin{centering}
\begin{tabular}{ccc}
\hline
\hline
\multirow{2}{*}{Configuration} & \multicolumn{2}{c}{Carbon sites }\tabularnewline
\cline{2-3}
 & Reaction 1, 2, 3 & Reaction 4\tabularnewline
\hline
A & \textbf{1}, 2, 3 & \textipa{\LyXbar{}}\tabularnewline
B & \textbf{1}, 3, 4 & 1, 2, 3, 4\tabularnewline
C & \textbf{1}, 4, 5 & 1, 3, 4, 5\tabularnewline
D & \textbf{1}, 5, 6 & \tabularnewline
E & \textbf{1}, 6, 7 & 1, 5, 6, 7\tabularnewline
F & \textbf{1}, 7, 8 & \textipa{\LyXbar{}}\tabularnewline
G & \textbf{1}, 8, 9 & 1, 7, 8, 9\tabularnewline
H & 1, 3, 4, 5, 6 & \textipa{\LyXbar{}}\tabularnewline
I & 1, 5, 6, 7 & \textipa{\LyXbar{}}\tabularnewline
K & \textipa{\LyXbar{}} & 1, 3, 5, 6\tabularnewline
L & \textipa{\LyXbar{}} & 1, 5, 6, 8\tabularnewline
\hline
\hline
\end{tabular}
\par\end{centering}

\end{table}

\begin{figure}[h]
\begin{centering}
\includegraphics{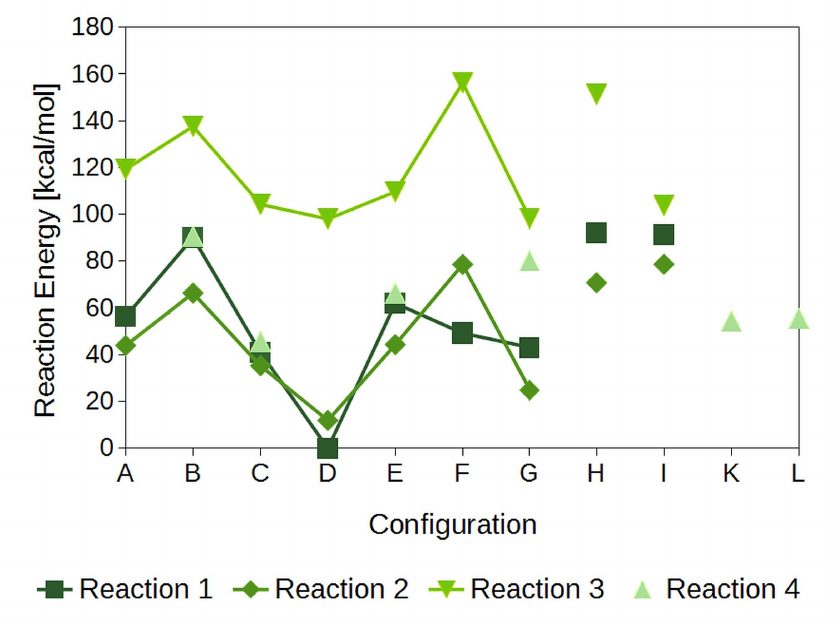}
\par\end{centering}

\caption{(Color online) Reaction energies of all studied pre-treatment reactions.
Dark green for reaction 1 (oxygen molecule), medium green for reaction
2 (oxygen and water molecules), light green for reaction 3 (2 water
molecules) and very light green for reaction 4 (oxygen and 2 water
molecules). For reactions 1-3 the symbols for configuration A-G are
connected due to the anti-clockwise movement of the functional groups
(cf. fig. \ref{fig:3}b and table \ref{tab:2}) \label{fig:4}}

\end{figure}

The calculated reaction energies for the different pre-treatment reactions
(cf. eq. \ref{eq:1}) are shown in fig. \ref{fig:4}. The figure illustrates
that the reaction energy depends on the type of the functional groups
as well as on their position at the SV defect. Using configuration
D1 (configuration D and reaction 1) as the reference, all other studied
functionalization reactions yield higher positive values of the reaction
energy. We can conclude that this reference reaction of the configuration
D1 is proven to be the most stable structure throughout all studied
reactions.

The dominating influence of the type of the functionalization reaction
on the reaction energy is clearly visible in fig. \ref{fig:4}. The
energies of reaction 3), involving water only, are significantly higher
than those of the reactions 1), 2) and 4) involving oxygen which all
yield similar reaction energies for most of the positions. The difference
of about 75 kcal/mol can be explained by comparing the sum of the
binding energies of all bonds which are broken and formed in the course
of these reactions. While the net bonding of reactions 1) and 2) and
4) are comparable and lie in the range of 220-230~kcal/mol,
it reaches only 154~kcal/mol for reaction 3) which is indeed by 66-76~kcal/mol
lower. We conclude that reactions involving oxygen are more likely
to occur and that the formation of C-H groups is unfavorable in a
mixed oxygen/water atmosphere.

Besides the influence on the type of functionalization reaction, fig.
\ref{fig:4} shows a clear dependence of the reaction energy on the
configuration of the involved sites. Surprisingly, the position dependence
of the reaction energy appears to be nearly identical for all four
reactions. It is clearly visible in fig. \ref{fig:4} that the position
dependence of the reaction energies reflects the mirror symmetry of
the single vacancy defect (cf. fig. \ref{fig:3}a). The most symmetric
configuration D is the lowest in energy for the reactions 1), 2) and
3). It is characterized by double bonds at the carbon position 1 and
two single bonds situated at the opposite carbon positions. Around
the defect structure, the configurations A/G, B/F and C/E should be
identical which is true for our data within the typical accuracy range
of the DFT results.

Besides the systematic variation of the carbon positions around the
defect structure, a few other configurations have been studied which
are also shown in fig.\ref{fig:4}. These configuration were chosen
in accordance to the low energy of D1, where the oxygen atom not only
supplies the missing electrons, but also re-completes one hexagon
ring of the defect (see in fig. \ref{fig:3}b). Based on the configurations
H and I, we can conclude that the double bond at the carbon atom 1
is of crucial importance for the reaction energy (cf. fig. \ref{fig:3}a).
Reaction 1) and 2) allow for such a double bond to be formed at atom
1. However, due to steric hindrance it is impossible for two hydroxyl
groups to bind to carbon atom 1 in reaction 4). Surprisingly this
does not significantly raise the overall reaction energy of reaction
4).

In contrast, replacing the double bonded oxygen atom at position 1
with one hydrogen atom and one hydroxyl group in case of reaction
3) increases the reaction energy. This is the effect of the repulsion
between the hydrogen atom and hydroxyl group at the carbon atom 1.
Configuration I illustrates this clearly. Opposed to A-F, in I1, I2
and I3 no double bonded oxygen atom is present at carbon atom 1, bringing
the energies for reaction I1, I2 and I3 closer together than for any
other configuration.

In conclusion, our pre-treatment study shows that the carbon atom,
where the functional group is bound to, influences the reaction energy.
This conclusion results in the fact that structure D1 is the most
stable one, as the carbon atom 1 is double bonded to an oxygen atom.

The next step is to either investigate whether it
is possible to perform the TMA-ALD without the presence of hydroxyl
groups similar to some exceptional cases on metal surfaces \citep{lu2014first,masango2016probing}
or to find a way to hydroxylate the CNT. The first possibility is briefly
explained in the next section.

In the latter case this can be achieved either by finding an energetic
more favorable structures saturated by hydroxyl groups or by finding
a reaction path that turns this oxygen defect into a hydroxylated
one. Since the first approach is not straight forward and includes
the investigation of a wide variety of structures, we decided to follow
the latter option.

\subsection{TMA ALD Process Performed on CNTs\label{sub:TMA-ALD-Process}}

\begin{figure}[h]
\begin{centering}
\includegraphics{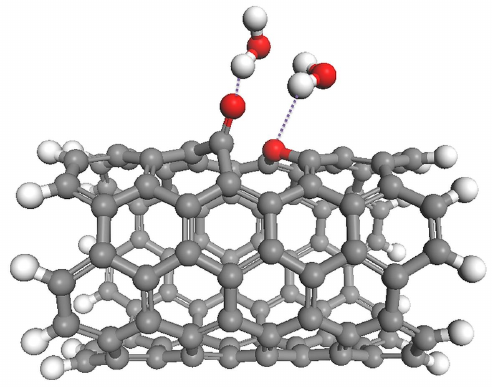}
\par\end{centering}

\caption{(Color online) Two water molecules bound to the dominant oxygen structure
D1, forming D1$_{\textrm{H}_{2}\textrm{O}}$.\label{fig:5}}

\end{figure}

In order to obtain the required hydroxyl groups for the initial TMA
ALD step, we let the dominant pre-treated CNT configuration D1 react
with water. In our simulations the water molecules are forced to react
with the functional groups and not with the CNT itself. Under these
conditions, we are able to bond water molecules via hydrogen bonds
to the oxygen atom of the D1 structure (see fig. \ref{fig:5}). This
structure is further referred to as D1$_{\textrm{H}_{2}\textrm{O}}$.

The investigated reactions show that the bonding of the water molecules
to the oxygen atom is energetically favorable by 5 kcal/mol per bonded
water molecule. In the resulting structure the water molecules are
close enough to the SV defect of the CNT such that they can function
as hydrogen donors for the reaction with TMA.

For the course of the ALD reactions the chemisorption of TMA has to
be checked. As a concurrent reaction the water may react with TMA
prior to its adsorption on the oxygen-functionalized defect D1. Thus
we investigate whether the TMA molecule, as desired, is able to bond
to D1$_{\textrm{H}_{2}\textrm{O}}$.

\begin{figure}[h]
\includegraphics{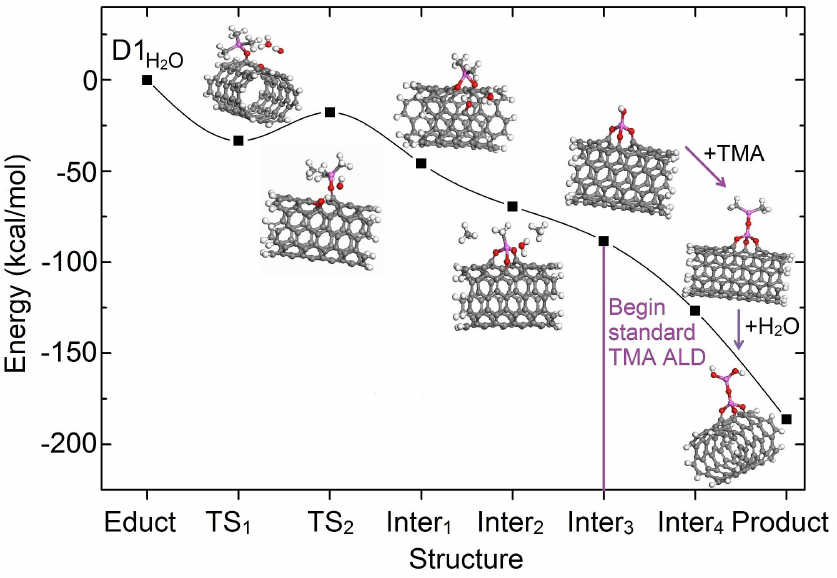}

\caption{(Color online) The alternative reaction pathway for the TMA ALD process.
Beginning with the educt D1$_{\textrm{H}_{2}\textrm{O}}$ two TMA
molecules react one after another with the CNT, forming transition
states (TS) and intermediate states (Inter) until the final product
is formed. The purple line indicates where the standard TMA ALD process
starts. first with the TMA addition, followed by the reaction of water.
\label{fig:6}}
\end{figure}

In fig. \ref{fig:6} we show the detailed reaction pathway for the
first and second TMA molecule reacting with D1$_{\textrm{H}_{2}\textrm{O}}$.
The zero point reference energy for this reaction part is the educt
D1$_{\textrm{H}_{2}\textrm{O}}$ (cf. fig. \ref{fig:5}). All reaction
energies are calculated with reference to this system. Thus, the reaction
energy $E_{R}$ is defined as:

\begin{equation}
E_{R}=E_{n\:CNT}+m*E_{CH_{4}}-E_{(n-1)\:CNT}-E_{TMA},\label{eq:2}
\end{equation}
with $E_{n\:CNT}$ being the energy of the CNT after the $n$-th ALD
sub-step, $E_{(n-1)\:CNT}$ the energy of the CNT before the $n$-th
ALD sub-step, $m$ is the number of the created methyl groups and
$E_{TMA}$ the energy of the TMA-precursor.

Fig. \ref{fig:6} shows that in the first reaction step the TMA molecule
reacts with the oxygen atom that shares a double bond with carbon
atom 1. The replacement of the C=O bond with a C-O-Al bond to create
TS$_{1}$ (transition state 1) is exothermic by 33.3~kcal/mol.
In a next step, one of the two water molecules donates one hydrogen
atom to form the first CH$_{4}$ molecule. The remaining hydroxyl
group of the water molecule bonds with carbon atom 6, forming TS$_{2}$.
The reaction barrier for this step (cf. fig. \ref{fig:6}) is 15.6
kcal/mol. This barrier height is in good agreement with the literature
value of about 12~kcal/mol \citep{ALD_George}.

We also investigated the reaction of TMA with the non-hydroxylated CNT D1,
which is presented in detail in the appendix. We found that while the
reaction energy for the first transition state TS$_{1}$ is similar between
D1$_{\textrm{H}_{2}\textrm{O}}$ and D1, the further reaction steps for the
non-hydroxylated CNT D1 are less favorable.

During the following two reaction steps the aluminum atom is further
anchored to the middle of the SV defect. From TS$_{2}$ to Inter$_{1}$
(intermediate state 1) the oxygen atom of the initial D1 turns towards
the aluminum atom and the newly introduced hydroxyl group reacts with
methyl to release a second CH$_{4}$ molecule (cf. Inter$_{2}$ in
fig. \ref{fig:6}).

The reaction of the second water molecule with the precursor results
into the structure Inter$_{3}$, which completes the reaction of the
first TMA molecule with the CNT at -88.5~kcal/mol.
The resulting structure obtains one hydroxyl group, which is the seed
for a standard TMA ALD cycle. The new cycle does not rely on the presence
of nearby water molecules anymore. According to \citep{TMA_Energy}
the addition of one TMA molecule is an exothermic reaction by 39~kcal/mol.
This is in agreement with our value of the reaction energy for the
addition of the second TMA molecule (38.3~kcal/mol). A similar agreement
is found for the second ALD half cycle where two water molecules from
the gas phase react with Inter$_{4}$ to the final product of our
investigation. For this step literature \citep{TMA_Energy} predicts
an energy of 68~kcal/mol, which is in the same range as our DFT reaction
energy of 58.8~kcal/mol.

We can conclude that it is possible to bond one aluminum atom to the
oxidized CNT defect by ALD via adsorbed water molecules, which form
hydrogen bonds to the oxygen atoms. Once this first TMA molecule is
bound to the CNT, a seed is provided from where a standard TMA ALD
process can be performed.

\section{Conclusions }

We present a study on the reaction of a metallic, non-periodic (5,5)-CNT
with a SV defect in a wet oxygen atmosphere. We show that the reaction
energy of a number of pre-functionalization reactions symmetrically
depends on the position of the functional groups, wherein carbon atom
1 takes a special position in the symmetry axis (cf. fig. \ref{fig:3}a).
While a saturation of the SV defect with 4 hydroxyl groups would be
ideal, we demonstrate that a defective CNT preferably reacts with
oxygen.

In order to functionalize such an oxidized CNT via TMA ALD, water
molecules have to be adsorbed via hydrogen bonds to the oxygen atoms
of the defect. These water molecules act as the hydrogen donors for
the first TMA ALD step. Once this step is successfully completed,
a standard TMA ALD cycle can be performed on the CNT with hydroxyl
groups on the first deposited aluminum atom acting as the seed.

As the basis for this alternative TMA ALD process relies on the SV
defect of the CNT, we can generalize this process for other metallic
and semiconducting CNTs containing SV defects. This method could also
be used as an initial step for other ALD processes that rely on hydroxyl
groups, but where only oxygen is present on the surface.

\clearpage{}\bibliographystyle{elsarticle-num}

\clearpage{}

\appendix
\setcounter{figure}{0}

\counterwithin{figure}{section}

\section{Supporting information}

\subsection{Position of functional groups}

The following figures \ref{fig:Si_1} to \ref{fig:Si_3} show
the position of the functional groups on the SV defect for the reactions
2)-4).

\begin{figure}[h]
\begin{raggedright}
\includegraphics{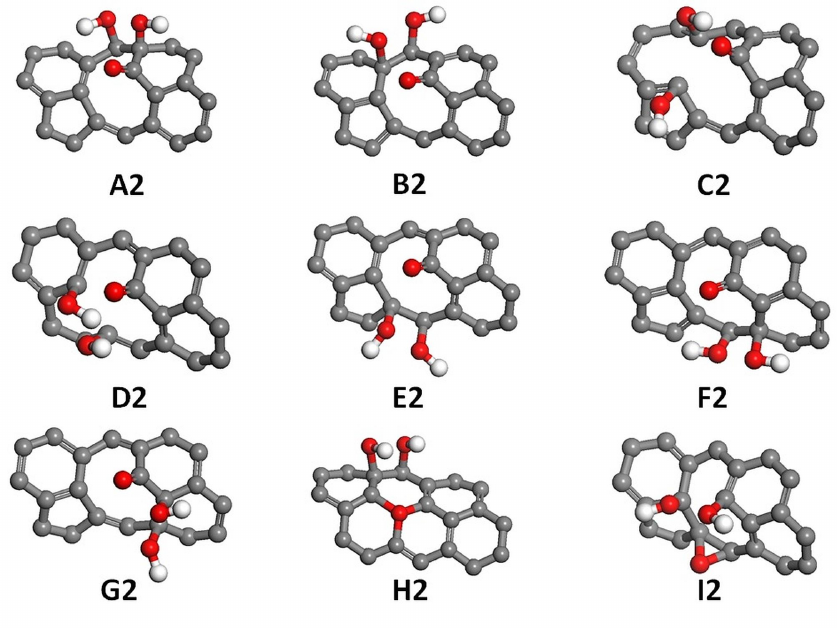}
\par\end{raggedright}

\caption{(Color online) Positions of the functional groups for Reaction 2)
with oxygen and one water molecule. \label{fig:Si_1} }
\end{figure}

\begin{figure}[h]
\begin{raggedright}
\includegraphics{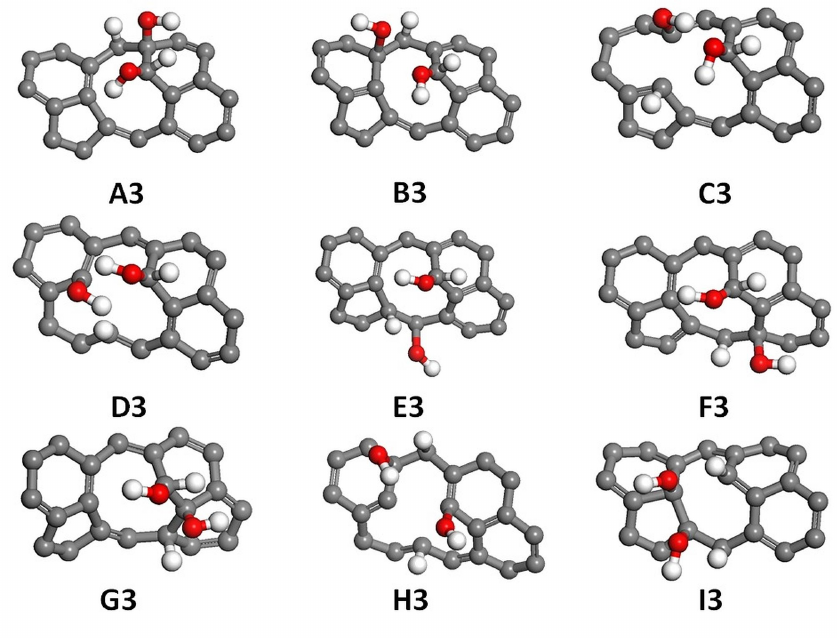}
\par\end{raggedright}

\caption{(Color online) Positions of the functional groups for Reaction 3)
with 2 water molecules. \label{fig:Si_2} }
\end{figure}

\begin{figure}[h]
\includegraphics{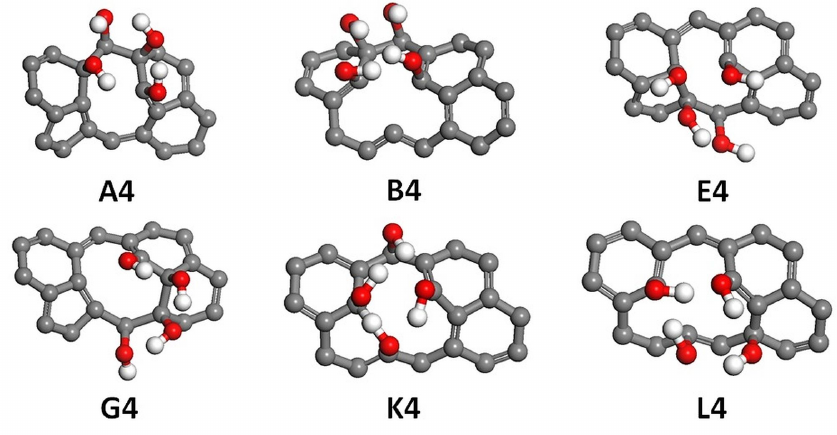}

\caption{(Color online) Positions of the functional groups for Reaction 4)
with oxygen and 2 water molecules. \label{fig:Si_3} }
\end{figure}

\subsection{{TMA ALD on non-hydroxylated CNTs}}

\begin{figure}[h]
\includegraphics{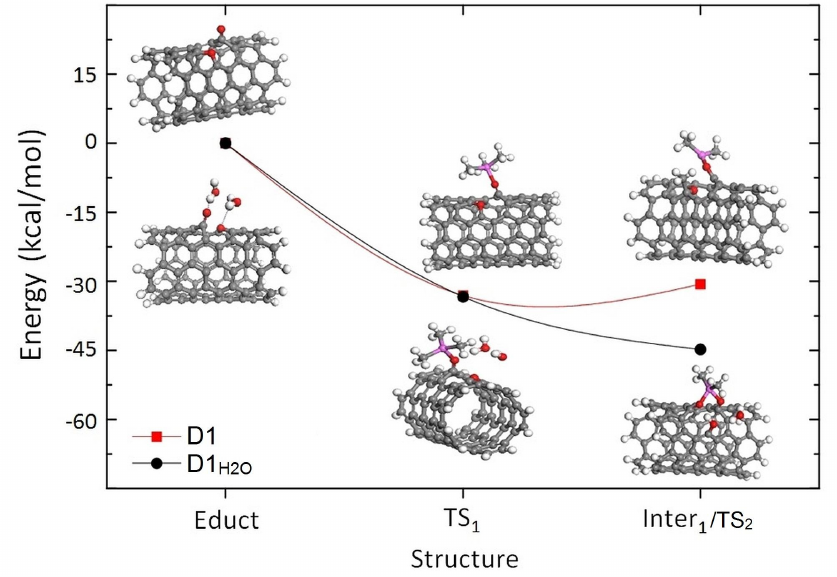}

\caption{(Color online) Comparison of the first TMA ALD reaction step on the
structure D1 (red) and D1$_{\textrm{H}_{2}\textrm{O}}$ (black). The
pictures in the top half show the structure for the reaction with
D1 and the structures in the bottom half belong to the reaction with
D1$_{\textrm{H}_{2}\textrm{O}}$. \label{fig:SI_4}
}
\end{figure}

Here, we want to briefly discuss the TMA ALD
process that is performed on the dominant pre-treated CNT configuration
D1 without the presence of water molecules. Figure \ref{fig:SI_4}
shows a comparison of the first initial step between the configuration
D1 and the hydroxylated structure D1$_{\textrm{H}_{2}\textrm{O}}$
as discussed in the sub-section \ref{sub:TMA-ALD-Process}. It is visible that while the transition states of D1 and D1$_{\textrm{H}_{2}\textrm{O}}$
are similar in energy ( -33.1~kcal/mol and -33.3~kcal/mol, respectively) the energy of the next reaction step differs.

The intermediate state Inter$_1$ from the starting structure D1$_{\textrm{H}_{2}\textrm{O}}$
is more stable (-44.8~kcal/mol) than the transition state TS$_2$
of D1 (-30.6~kcal/mol). This makes the reaction
with the starting structure D1$_{\textrm{H}_{2}\textrm{O}}$ favorable. We note that
'analogous to the pre-treatment' the methyl group has many possible bonding sites, with the carbon atoms
1-9 of the SV defect being the most likely candidates. The structure as shown on figure \ref{fig:SI_4} is the
reaction site which so far is the most promising. Further studies would be neceesary to affirm this preliminary finding.

\begin{figure}[h]
\includegraphics{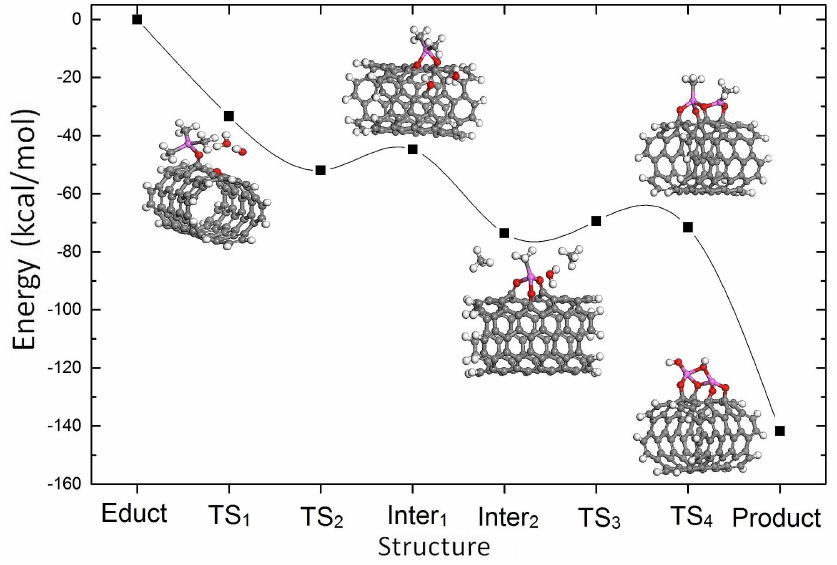}

\caption{(Color online) An alternative TMA ALD reaction of the educt D1$_{\textrm{H}_{2}\textrm{O}}$
where 2 TMA molecules bind directly to the SV defect. \label{fig:Si_5}
}

\end{figure}

The methyl group that binds to the
CNT at D1 takes away one possible anchor site for the TMA molecule, which makes the non-hydroxylated pathway less favorable.
This can prevent the TMA molecule from forming a more stable bond/connection
to the CNT. In the case of D1$_{\textrm{H}_{2}\textrm{O}}$, on the
other hand, it is possible to anchor up to two TMA molecules at the
SV defect of the CNT (cf. fig. \ref{fig:Si_5}).

\end{document}